\documentclass[aip,cha,superscriptaddress,numerical,preprint]{revtex4-1}

\usepackage{amssymb,amsmath}
\usepackage{graphicx} % Include figure files
\usepackage{epsfig}
\usepackage{bm}
\usepackage{color}
\usepackage[normalem]{ulem}

\newcommand\beq{\begin{equation}}
\newcommand\eeq{\end{equation}}
\newcommand\beqa{\begin{eqnarray}}
\newcommand\eeqa{\end{eqnarray}}

\newcommand{\nn}{\nonumber \\}

\begin{document}

\title{Are Continuum Predictions of Clustering Chaotic?}

%\date{\today}

\author{William D. Fullmer}
\email[E-mail: ]{william.fullmer@colorado.edu}
\affiliation{Department of Chemical and Biological Engineering, University of Colorado, Boulder, CO 80309, USA}

\author{Christine M. Hrenya}
\email[Corresponding Author, E-mail: ]{hrenya@colorado.edu}
\affiliation{Department of Chemical and Biological Engineering, University of Colorado, Boulder, CO 80309, USA}

\begin{abstract}
Gas-solid multiphase flows are prone to develop an instability known as clustering. Two-fluid models, which treat the particulate phase as a continuum, are known to reproduce the qualitative features of this instability, producing highly-dynamic, spatiotemporal patterns. However, it is unknown whether such simulations are truly aperiodic or a type of complex periodic behavior. By showing that the system possesses a sensitive dependence on initial conditions and a positive largest Lyapunov exponent, $\lambda_1 \approx 1/\tau$, we provide a tentative answer: continuum predictions of clustering are chaotic. We further demonstrate that the chaotic behavior is dimensionally dependent, a conclusion which unifies previous results and strongly suggests that the chaotic behavior is not a result of the fundamental kinematic instability, but of the secondary (inherently multidimensional) instability.  
\end{abstract}

%\pacs{05.20.Dd, 45.70.Mg, 51.10.+y, 47.50.+d, 81.05.Rm, 47.11.-j}
%\keywords{granular media, viscous dissipation; multiphase and particle-laden flows}

%\draft
\date{\today}
\maketitle

\begin{quotation}
Granular matter is a collection of discrete, interacting solid particles which, like classic (molecular) matter, can be generally classified into one of three states \cite{jaeger96}: i) under static conditions granular heaps or piles can sustain gravity-induced stress and behave like a solid \cite{geng01}; ii) dense granular flows characterized by enduring and multi-particle contacts behave similarly to a fluid \cite{forterre08}; and iii) rapid granular flows characterized by instantaneous contacts described as a granular gas \cite{goldhirsch03}. However, it is worth noting that all three granular states are only superficially similar to their molecular counterparts due to the dissipative nature of particle-particle contacts (inelasticity, friction, etc.) \cite{jaeger96, goldhirsch03}. In industrial operations, “rapid” (collision-dominated) granular flows are often encountered in devices in which the particles are fluidized by a gas \cite{kunii}; such gas-solid flows are the focus of this work. Rapid gas-solid flows are prone to an instability termed clustering in which particles tend to form spatially inhomogeneous patterns of high and low concentrations \cite{fullmer17}.  Continuum or two-fluid models have long been known to be able to predict the qualitative nature of the clustering instability \cite{agrawal01} and more recent quantitative assessments have also shown promising results \cite{mitrano14, fullmer16, fullmerJFM} However, it is yet unknown whether or not such predictions are chaotic. We take a first step in answering this question by simulating fluidization in an unbounded domain and calculating a positive largest Lyapunov exponent, thereby indicating that continuum predictions of clustering are in fact chaotic. Further, we show that chaotic behavior may be reduced to periodic behavior by constraining the dimensionality of the system. 
\end{quotation}

\section{Introduction}
\label{sec.intro}
Physicists and engineers have long been attracted to granular matter due to the rich variety of patterns and complex behavior they are prone to display \cite{hopkins91, goldhirsch93, goldfarb02, conway04, aranson06, vinningland07, cheng08, christov10, seiden11, shinbrot15}. Granular and multiphase gas-solid flows are also of central importance to energy, chemical, petrochemical, pharmaceutical, food processing and other key industrial sectors \cite{kunii, cocco14}. In this work, we are interested in rapid gas-solid flows, a regime characterized by nearly instantaneous and binary collisions, large Stokes numbers, $St \sim \rho_s U d_p / \mu_g$, and relatively dilute solids concentrations, which commonly occur in the freeboards or risers of circulating fluidized beds, pneumatic conveying systems, etc.

Industrial particulate systems are often characterized by a large separation of scales \cite{vanderhoef08, li13, cocco17} necessitating the use of continuum or two-fluid models (TFM) to make full, system-scale predictions in most cases. Unlike molecular dynamics (MD) or discrete element methods (DEM), which track the motion and collisions of every particle, continuum models only resolve average properties, e.g., solids volume concentration, mean velocity and granular temperature -- a measure of the fluctuating kinetic energy of the solids phase. An analogy between particles and molecules is often employed using a kinetic theory (KT) approach to derive and constitute continuum models \cite{gidaspow, brilliantov, pannala}. We use the nomenclature KT-TFM here to indicate the multiphase nature of the continuum model, rather than the more common KTGF (kinetic theory of granular flows), which is often used interchangeably for granular flows (no interstitial phase) and multiphase flows.

The behavior of interest in rapid gas-solid flow is the clustering instability \cite{fullmer17}. Through the dissipation of granular energy and/or a mean relative motion between the phases, initially uniform distributions of particles tend to group together to form inhomogeneous distributions of dense clusters which are persistent yet dynamic in nature. With sufficient grid resolution, KT-TFM simulations are able to qualitatively predict the clustering behavior \cite{agrawal01}. A growing body of work also suggests that continuum model predictions are also quantitatively accurate in the prediction of clusters \cite{mitrano14, fullmer16, fullmerJFM}.

It is reasonable to expect that direct numerical simulations (DNS) \cite{yin13} and CFD-DEM simulations \cite{radl14, capecelatro15}, which resolve the particle-scale dynamics, should be chaotic as elastic  \cite{dellago97} and inelastic \cite{mcnamara01} hard-sphere MD simulations are known be chaotic, even in the absence of clustering. For continuum predictions, however, the picture is not so clear.  Several observations suggest that continuum predictions of clustering may be chaotic: i) visual inspection of the instantaneous patterns do not appear regular enough to be periodic \cite{agrawal01, fullmer16}; ii) continuum simulations of a wall bounded liquid-solid fluidized bed have been shown to exhibit a continuous solids concentration spectra \cite{gevrin10}; and iii) experimental measurements of gas-solid flows in risers are known to be chaotic \cite{marzocchella97, ji00} (although, clearly, physical gas-solid flows inherently discrete). Other observations, however, appear to indicate that continuum predictions of clustering are not chaotic: i) a previous study of clustering in a dilute riser flow (wall bounded fluidization) with a KT-TFM only revealed periodic behavior \cite{benyahia07}; ii) it is possible that chaos in such simulations may only arise due to the on/off and/or very stiff (highly-nonlinear) behavior of the empirical models used to close the solid stresses near maximum packing \cite{benyahia16}; and iii) when caused by mean relative motion, the clustering instability in gas-solid flow is known to stem from a kinematic instability \cite{fullmer17}, a type of instability which is known to produce limit cycle behavior in gas-liquid fixed-flux TFMs \cite{robinson08, bertodano}.

Due to the competing suggestions noted above and in the absence of definitive evidence, it is unknown whether continuum predictions of clustering are chaotic or some type of complex limit cycle, e.g., a high-order n-torus. We seek to answer this outstanding question by studying KT-TFM predictions of clustering in gas-solid fluidization in a fully-periodic domain. We first study the nonlinear response of the system to small changes in the initial condition and then calculate the largest Lyapunov exponent from a pair of coupled simulations. Finally, we conclude by showing how the dimensionality of the system can affect the observed behavior, thereby resolving the aforementioned discrepancies.

\section{Two-Fluid Model}
\label{sec.tfm}
The KT-TFM used in this work is one of the most rigorous of such models to-date owing to the instantaneous particle-gas force included directly in the starting Enskog equation \cite{garzo12}. For the sake of brevity, only the transport equations are included here: 
\beq
\frac{{\partial \phi }}{{\partial t}} + \nabla  \cdot \phi {{\bf{U}}_s} = 0 ,
\label{eq.mass.s}
\eeq
\beq
\frac{{\partial (1 - \phi )}}{{\partial t}} + \nabla  \cdot (1 - \phi ){{\bf{U}}_g} = 0 ,
\label{eq.mass.g}
\eeq
\beqa
{\rho _s}\phi \left( {\frac{{\partial {{\bf{U}}_s}}}{{\partial t}} + {{\bf{U}}_s} \cdot \nabla {{\bf{U}}_s}} \right) =  - \phi \nabla {p_g} - \nabla {p_s} \nonumber \\
+ \nabla  \cdot {{\bf{\sigma }}_s} - \beta \left( {{{\bf{U}}_s} - {{\bf{U}}_g}} \right) + {\rho _s}\phi {\bf{g}}
\label{eq.mom.s}
\eeqa
\beqa
{\rho _g}(1 - \phi ) \left( {\frac{{\partial {{\bf{U}}_g}}}{{\partial t}} + {{\bf{U}}_g} \cdot \nabla {{\bf{U}}_g}} \right) =  - (1 - \phi )\nabla {p_g} \nonumber \\
+ (1 - \phi )\nabla  \cdot {{\bf{\sigma }}_g} + \beta \left( {{{\bf{U}}_s} - {{\bf{U}}_g}} \right) + {\rho _g}(1 - \phi ){\bf{g}}
\label{eq.mom.g}
\eeqa
\beqa
\frac{3}{2}{\rho _s}\phi \left( {\frac{{\partial T}}{{\partial t}} + {{\bf{U}}_s} \cdot \nabla T} \right) = \left( {{{\bf{\sigma }}_s} - {p_s}{\bf{I}}_3} \right):\nabla {{\bf{U}}_s} - \nabla  \cdot {\bf{q}} \nonumber \\
- \frac{3}{2}{\rho _s}\phi T{\zeta _1}\left( {\nabla  \cdot {{\bf{U}}_s}} \right) + \frac{3}{2}{\rho _s}\phi \left( {\xi  - \frac{{2\gamma }}{m}T - {\zeta _0}T} \right). \quad
\label{eq.T}
\eeqa
The nine unknown variables to be solved for are the solids (volumetric) concentration, $\phi$, the solids- and gas-phase velocity vectors, ${\bf{U}}_s$ and ${\bf{U}}_f$, the gas pressure, $p_g$, and the granular temperature, $T$, a measure of the (isotropic) fluctuating kinetic energy in the solids phase. Solids-phase constitutive relations for the stress tensor, ${\bf{\sigma }}_s$, and granular heat flux, ${\bf{q}}$, and closures for the pressure, $p_s$, shear and bulk viscosity, granular conductivity, Dufour coefficient, and the first- and zeroth-order collisional cooling rates, $\zeta_1$, and $\zeta _0$, are derived from kinetic theory and can be found in the original work \cite{garzo12}. Gas-solid interaction closures for the mean drag \cite{beetstra07}, $\beta$, the thermal drag, $\gamma$ and the neighbor effect $\xi$ are all derived from DNS data. (We note that the first-order thermal-Reynolds-number dependent term of the thermal drag model was re-fit to the original DNS data \cite{wylie03} with a function that vanished in the zero concentration limit, $K ( \phi ) = \sqrt{ 0.3 \phi } / (1 - \phi)^{3.6}$.) The gas-phase shear viscosity, $\mu_g$, density, $\rho_g$ and the density of the constituent particles, $\rho_s$, are material properties and assumed constant, i.e., the gas-phase is treated as incompressible. The gravitational acceleration vector, ${\bf{g}}$ is also assumed constant and aligned in the vertical, $y$-dimension. Finally, it should be pointed out that although the instantaneous fluid-particle force is modeled with a Langevin equation \cite{garzo12}, once the Enskog equation is averaged over the velocity distribution function the resulting KT-TFM is entirely deterministic.

The KT-TFM described above is solved numerically using the National Energy Technology Laboratory’s open-source MFiX code (\url{https://mfix.netl.doe.gov/}). MFiX is a finite-volume-based CFD code which solves the governing equations on a staggered grid with implicit Euler time advancement using a SIMPLE-type algorithm with variable time stepping. The Superbee flux-limiter is applied for all variable extrapolation.

The KT-TFM and the solution method described above, as well as the system and conditions described in Sec.~\ref{sec.sys}, are the same as those from a previous work where the primary focus was on validating the mean slip velocity against CFD-DEM data \cite{fullmer16}. However, several specific changes have been made to eliminate "artificial" potential sources of chaos. No empirical solids (frictional) stress model is considered; even the pressure-like term used to limit the solids concentration below the maximum packing limit has been neglected here. Consequently, the radial distribution function (RDF) at contact of Carnahan and Starling \cite{carnahan69} is used in place of the Ma and Ahmadi \cite{ma88} RDF. While there is little quantitative difference between the two RDFs for $\phi < 0.5$, the latter does not have the appropriate asymptotic behavior as $\phi \to \phi_{\max} \approx 0.64$, only diverging in the unphysical limit of $\phi \to 1$. In the previous study \cite{fullmer16} a dynamic flux renormalization, $j_y = \phi v_s + (1 - \phi) v_g = 0$, was applied iteratively to ensure that the frame of reference did not drift due to numerical precision, i.e., round-off errors . Except where noted, the renormalization has been removed here so that $j_y$ can, and does, drift from zero. Finally, the simulations reported here have been run in serial. Although serial computation increased the computational demands considerably (previous simulations \cite{fullmer16} were parallelized over 36 cores), serial simulations remove suspicion that chaos may be attributed to domain decomposition errors.

\section{System and Conditions}
\label{sec.sys}
Four nondimensional parameters determine the condition of the system: the Archimedes number, $Ar = \rho_g (\rho_s - \rho_g) \left| {\bf{g}} \right| d_p^3 / \mu_g^2$ where $d_p$ is the particle diameter, the density ratio $\rho^* = \rho_s / \rho_g$, the mean solids concentration $\left\langle \phi \right\rangle$ where the angle brackets indicate a volume average, and the restitution coefficient, $e$. While not appearing directly in Eqs.~(\ref{eq.mass.s}) - (\ref{eq.T}), the restitution coefficient describes the dissipation of energy due to particle collisions and appears in all of the solids-phase closures. In this KT-TFM \cite{garzo12}, $e$ is treated as a constant material property and takes a value of $e = 0.9$ here. The dimensional properties originally \cite{radl14} selected to represent typical fluid catalytic cracking particles in air correspond to $Ar = 24.886$ and $\rho^* = 1153.8$. Of the six mean concentrations considered previously \cite{fullmer16}, only the lowest, $\left\langle \phi \right\rangle = 0.02$, is considered here to avoid unphysical predictions near maximum packing.

The gas-phase pressure is decomposed into a local fluctuating component and a constant linear component that balances the force of gravity. The domain size is $L_x^* = L_z^* = 106.\bar 6$ in the transverse directions and four times as tall in the vertical (streamwise) direction, $L_y^* = 4 L_x^*$. The simulations specify a relatively fine, uniform grid of $N_x \times N_y \times N_z = 4 \cdot 32^3$. In the previous study, simulations were carried out from a randomly perturbed initial state to a final condition at $t^* = 76.8$. The final state is used as an initial condition in this work and the time is considered start at zero from this point in time. Here, space has been nondimensionalized by the particle diameter, $L^* = L/d_p$, and time is nondimensionalized by the viscous relaxation time, $t^* = t/\tau$, where $\tau = \rho_s d_p^2 / 18 \mu_g$ is the viscous relaxation time.

\begin{figure}
\begin{center}
{\includegraphics[height=8in]{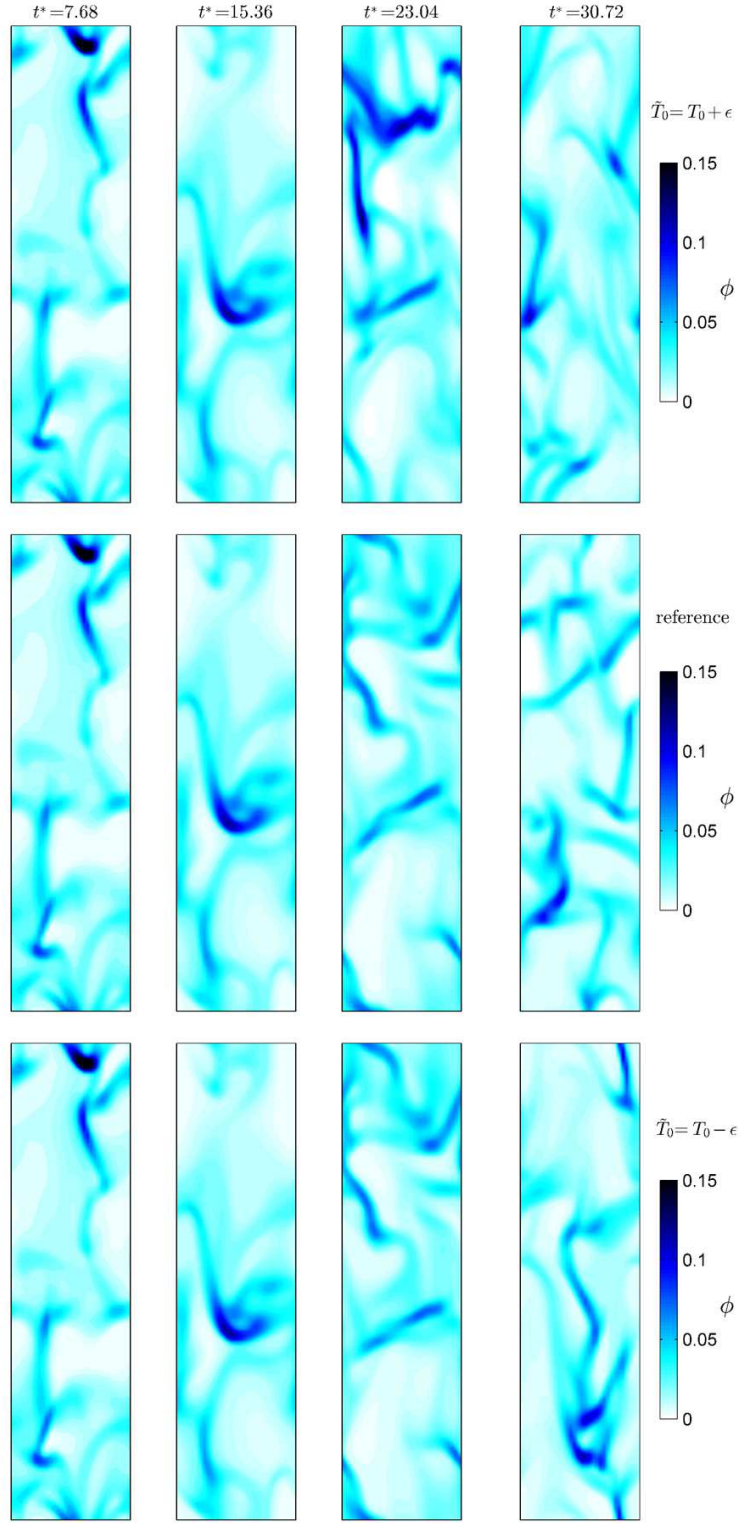}}
%{\includegraphics[height=2in]{Fig1a1.eps}}
%{\includegraphics[height=2in]{Fig1a2.eps}}
%{\includegraphics[height=2in]{Fig1a3.eps}}
%{\includegraphics[height=2in]{Fig1a4.eps}} \\
%{\includegraphics[height=2in]{Fig1b1.eps}}
%{\includegraphics[height=2in]{Fig1b2.eps}}
%{\includegraphics[height=2in]{Fig1b3.eps}}
%{\includegraphics[height=2in]{Fig1b4.eps}} \\
%{\includegraphics[height=2in]{Fig1c1.eps}}
%{\includegraphics[height=2in]{Fig1c2.eps}}
%{\includegraphics[height=2in]{Fig1c3.eps}}
%{\includegraphics[height=2in]{Fig1c4.eps}}
\end{center}
\caption{(Color online) Sensitivity to initial conditions: evolution of a reference trajectory and two perturbed cases with slightly higher (top) and lower (bottom) initial granular temperatures.}
\label{fig.1}
\end{figure}

\section{Sensitive Dependence on Initial Conditions}
\label{sec.sensic}
As a preliminary assessment of whether or not the system might be chaotic, we begin by studying the response of the system to slight changes in the initial condition. Three simulations were considered. The reference case is unperturbed, i.e., uses the final condition from a previous study. Two perturbed cases were considered with nearly the same initial condition but with the granular temperature uniformly perturbed by $T_0 \to T_0 \pm \epsilon$ where $\epsilon = 3.0 \cdot 10^{-5}$ is a constant, the magnitude of which if further justified in Sec.~\ref{sec.lle}, and $T_0$ is the unperturbed initial granular temperature field. We choose to perturb only the granular temperature in an effort to disturb the system as innocuously as possible.

The evolution of solids concentration from the three slightly different initial states is displayed in Fig.~\ref{fig.1} starting at $t^*$ = 7.68 and separated by the same amount. At the first two times (left half of Fig.~\ref{fig.1}), the three simulations are virtually indistinguishable. By $t^* = 23.04$,   the concentration in the top half of the domain of the perturbed case ${\tilde T_0} = T_0 + \epsilon$ (top row) is clearly different than the reference case. At this time, the lower half of the domain remains relatively similar. The second perturbed case ${\tilde T_0} = T_0 - \epsilon$ (bottom row) only displays minor deviations from the reference case. However by the time $t^* = 30.72$, the three simulations are completely different from one another, resembling three entirely different realizations of the same flow. This simple demonstration shows that the KT-TFM (continuum) prediction of clustering in unbounded fluidization -- in the absence of switches, functions containing singularities or highly-nonlinear empirical stress models – possesses a sensitive dependence on initial conditions, a strong hint that the flow may be chaotic.

\section{Largest Lyapunov Exponent}
\label{sec.lle}
While no formal proof of chaos exists \cite{sprott}, the Lyapunov spectrum remains one of the most generally accepted metrics in identifying chaotic behavior. Due to the size of this system, $N = 32 \times 128 \times 32 \times 9$ coupled ODEs, and its associated computational cost, we are only interested here in calculating the largest Lyapunov exponent (LLE), $\lambda_1^*$. A positive valued LLE indicates a chaotic system and, if positive, its magnitude determines the rate at which predictability of the system is lost. The LLE is defined by \cite{cencini}
\beq
\lambda _1^* = \mathop {\lim }\limits_{{t^*} \to \infty } \mathop {\lim }\limits_{{\delta _0} \to 0} \frac{1}{{{t^*}}}\ln \frac{{\delta ({t^*})}}{{{\delta _0}}} ,
\label{eq.lle.def}
\eeq
where $\delta_0$ is the initial perturbation and $\delta(t)$ is the evolution of the perturbation with time. By relaxing the inner limit of Eq.~(\ref{eq.lle.def}), which is meant to ensure linearity, one can obtain a rough estimate of the LLE using $\delta(t)$ from the two perturbed cases in Fig.~\ref{fig.1}. For this $N$-dimensional system, $\delta(t)$ is the L$_2$-norm of the difference between the perturbed and reference trajectory summed over all primary variables, i.e.,  
\beq
\delta ({t^*}) = \sqrt {\sum\limits_{j = 1}^{{N_z}} {\sum\limits_{k = 1}^{{N_y}} {\sum\limits_{i = 1}^{{N_x}} {{{\left\| {{{{\bf{\tilde \alpha }}}_{i,j,k}}({t^*}) - {{\bf{\alpha }}_{i,j,k}}({t^*})} \right\|}^2}} } } } , 
\label{eq.delt}
\eeq
where the tilde denotes the perturbed trajectory and $\alpha$ is the vector of unknown (solution) variables, ${\bf \alpha} = (\phi, u_g, v_g, w_g, p_g, u_s, v_s, w_s, T)$.

\begin{figure}
\begin{center}
{\includegraphics[width=3in]{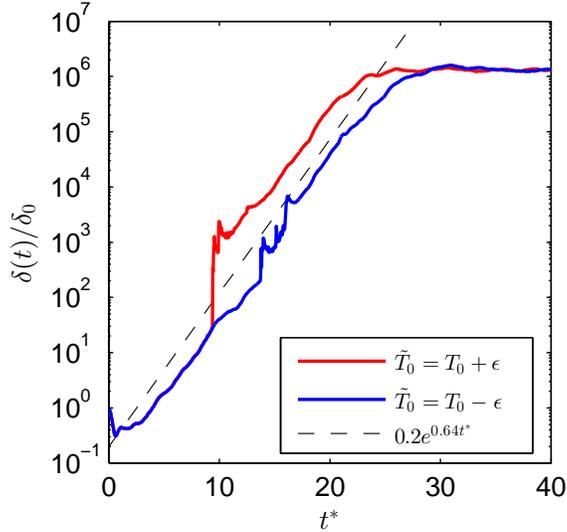}}
\end{center}
\caption{(Color online) Comparison of $\delta(t)$ for the two initially perturbed cases compared to the reference case. Divergence rate approximated by $e^{0.64 t ^*}$.}
\label{fig.2}
\end{figure}

The separation of the two initially perturbed cases relative to the reference case in Fig.~\ref{fig.1} are compared quantitatively in Fig.~\ref{fig.2}. After an initial period of contraction ($t^* < 1$), both cases diverge roughly exponentially from the reference case. At $t^* \approx 25$ and 30, both separations saturate when the three solutions have become essentially independent of one another, as shown in the fourth column of Fig.~\ref{fig.1}.  It is unknown what causes the abrupt changes, notably just before $t^* = 10$.  It was verified that this was not an output or post-processing error nor did either of the calculations need to reduce the time step for convergence around this time. The divergence can be fit to a rate of approximately $0.2 e^{0.64 t^*}$ shown as a dashed line in Fig.~\ref{fig.2}. Although this analysis is admittedly imprecise since the separation becomes too large to be considered in a linear regime, it nonetheless indicates that the LLE may be in the vicinity of $\lambda_1^* \sim 0.64$.

A more precise calculation of the LLE requires maintaining the inner linear-constraining limit in Eq.~(\ref{eq.lle.def}). In practice, the LLE is calculated by iteratively re-normalizing the perturbed trajectory to the reference or fiducial trajectory iteratively so that the two remain close \cite{wolf85, sprott, cencini}. Therefore, Eq.~(\ref{eq.lle.def}) becomes, 
\beq
\lambda _1^* = \mathop {\lim }\limits_{n \to \infty } \frac{1}{{n\Delta {t^*}}}\sum\limits_{i = 1}^n {\ln \frac{{{\delta _n}}}{{{\delta _0}}}} 
\label{eq.lle.calc}
\eeq
where $\delta_n$ is the expansion between times $t_{n-1}^*$ and $t_n^* = t_{n-1}^* + \Delta t^*$. After every $\Delta t^*$ step, the perturbed trajectory is re-normalized from the reference trajectory, 
\beq
{\bf{\tilde \alpha }}_{n+1} = {\bf{\alpha }}_n + ({\delta _0}/{\delta _n})({\bf{\tilde \alpha }}_n - {\bf{\alpha }}_n)
\label{eq.lle.renorm}
\eeq
preserving the magnitude of the initial separation, $\delta_0$ and the orientation of the previous step. Therefore the separation from $t_{n-1}^*$ to $t_n^*$ always begins with the same magnitude of initial separation, $\delta_0$. To carry out these calculations, two separate simulations were coupled together, periodically exporting solution data from the reference to the perturbed simulation. The time step used for LLE calculation $\Delta t^*$ is not the actual code iterative time step, $d t^*$. As discussed in Sec.~\ref{sec.tfm}, MFiX uses an adaptive time stepping to overcome non-convergence issues. It was determined that forcing the code time step to be small enough as to avoid non-convergence issues was prohibitively expensive for this calculation. Therefore, the code was allowed to freely adjust the time advancement step and was forced to sync to a constant $\Delta t^* = 3.84\cdot 10^{-3}$. Typically, each $\Delta t^*$ step contains between two to five iterative time advancement steps. As in Sec.~\ref{sec.sensic}, the reference case is restarted from the final state of previous calculations \cite{fullmer16} and the perturbed trajectory is initially offset by ${\tilde T_0} = T_0 - \epsilon$ with $\epsilon = 3.0 \cdot 10^{-5}$ for an initial separation of $\delta_0 = \sqrt{N} \epsilon / 3$. The value of $\epsilon$ is chosen so that when normalized over all $N$ ODEs, the separation is two orders of magnitude larger than the residual convergence criteria of the iterative-implicit Euler time advancement scheme, here set at 10$^{-7}$.

It takes approximately $t^* = 25$ for the initial perturbation to propagate throughout the system and align itself in the direction of maximum expansion, marked by a rather obvious change in the behavior of $\delta_n$. Perhaps not coincidentally, this time corresponds to the saturation of the initial perturbation in Fig.~\ref{fig.2}. We begin collecting data for the calculation of the LLE at $t^* = 38.4$ through the end of the simulation at $t^* = 76.8$. While this coupled-simulation pair took over 67 days to complete, this yields only $n = 10^4$ separations to average over, orders of magnitude smaller than other LLE measurements of spatio-temporal chaos \cite{brummitt09, fullmer14b} and not sufficient to report a converged value with significant precision. However, in this study we are less concerned with the exact value of $\lambda_1$ than its sign and, to a lesser extent, its approximate value. Therefore, we choose to split the data into ten equally spaced intervals. Of the ten samples, the minimum, average, and maximum values are calculated to be $\lambda_1^*$ = 0.6981, 0.9767, and 1.1714, respectively. Again, the most important result is that we have demonstrated with reasonable certainty that $\lambda_1^*$ is positive, indicating a mean divergence of the perturbed trajectory from the fiducial trajectory and signifying chaotic behavior. The rough (nonlinear) approximation from Fig.~\ref{fig.2} agrees relatively well with the $\lambda_1^*$ calculations, particularly the lower estimate. While approximate, the order of magnitude of $\lambda_1^*$ shows that the rate of predictability is lost inversely proportionally to the viscous relaxation time. It would be interesting to determine whether this finding holds in general, but is outside the scope of this exploratory study.

\begin{figure*}
\begin{center}
{\includegraphics[height=3in]{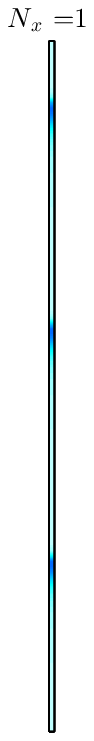}}
{\includegraphics[height=3in]{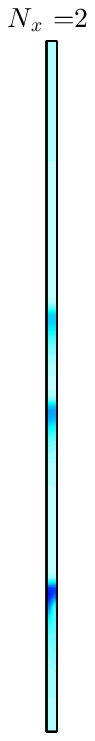}}
{\includegraphics[height=3in]{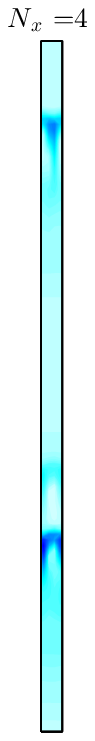}}
{\includegraphics[height=3in]{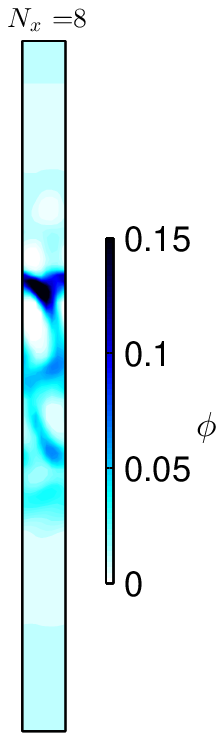}}
{\includegraphics[height=3in]{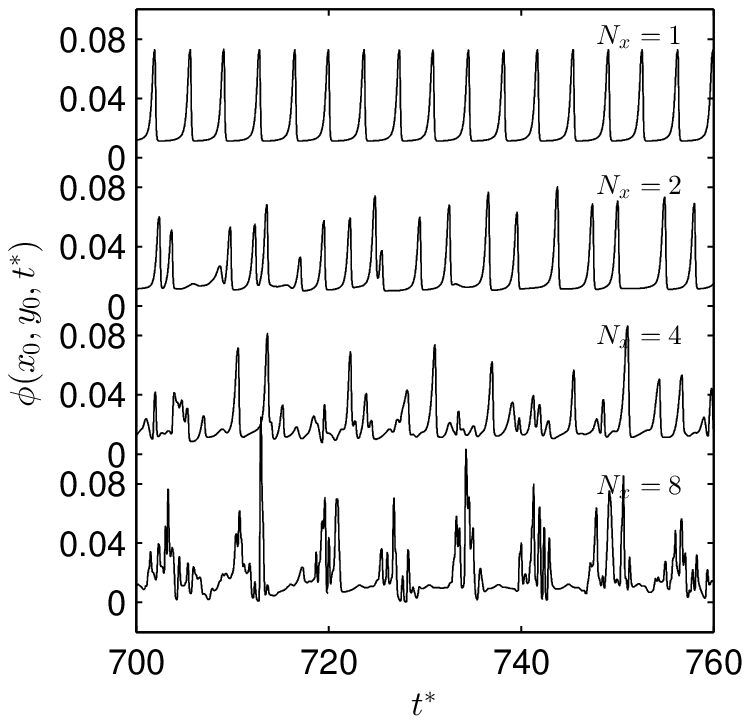}}
\end{center}
\caption{(Color online) Dimensionally dependent dynamics of clustering. On left, instantaneous contour plots for $N_x$ = 1, 2, 4 and 8 (left to right) transverse nodes. At right, the time-evolution of the solids concentration of the four cases at ($x_0$,$y_0$).}
\label{fig.3}
\end{figure*}

\section{Discussion}
\label{sec.close}
The simple, albeit computationally intensive, measurement of a positive LLE indicates that KT-TFM (continuum) predictions of clustering are chaotic in nature. While this seems to agree intuitively with the physical picture of the clustering instability, it begs the question of what is different about this case than previous results that reported only limit cycle behavior. The answer appears to lie in the dimensionality of the system studied here. The previously mentioned cases for particulate gas-solid \cite{benyahia07} and bubbly gas-liquid \cite{robinson08, bertodano} flows all considered a one-dimensional system. Here we provide a simple demonstration of the effect of dimensionality on clustering predictions by eliminating the $z$-dimensional dependence entirely and reducing the width of the system to $N_x$ = 1, 2 4 and 8 nodes. Note that the $N_x = 1$ case is purely one-dimensional. With width of each node is the same as in previous cases. These four simulations were initialized at rest with a random, normal perturbation. The flux renormalization procedure from the previous study \cite{fullmer16} was included so that limit cycle behavior could be identified without drift in the vertical flux. The reduced dimensionality simulations were run 10$\times$ longer than previously, up to times $t^* = 768$.

\begin{figure*}
\begin{center}
{\includegraphics[width=3.0in]{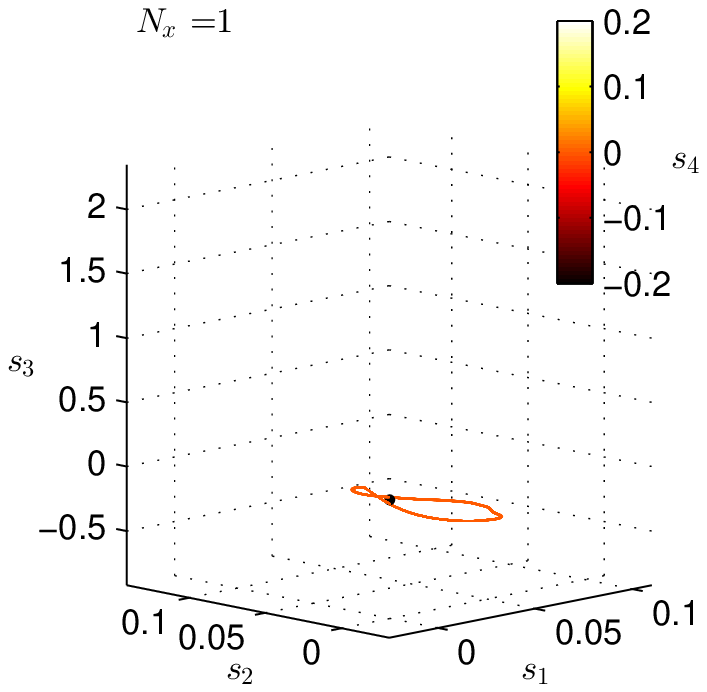}}
{\includegraphics[width=3.0in]{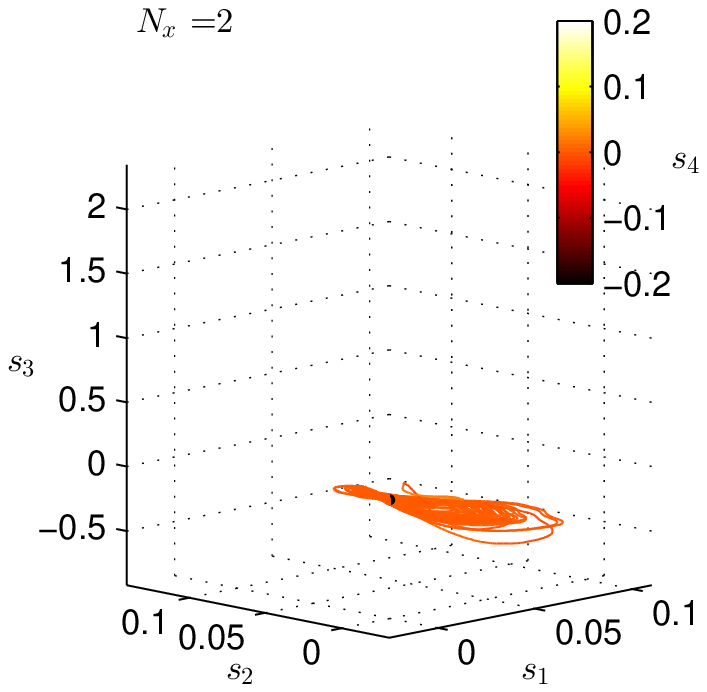}}
{\includegraphics[width=3.0in]{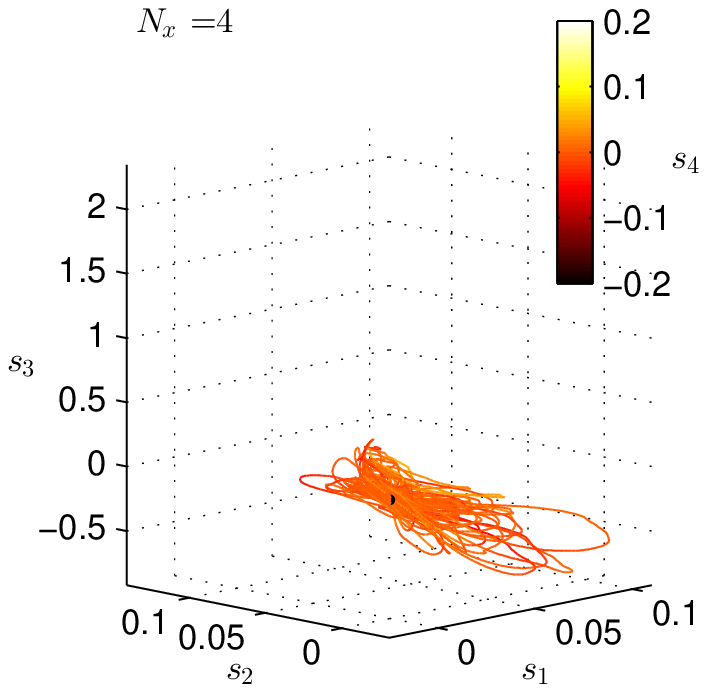}}
{\includegraphics[width=3.0in]{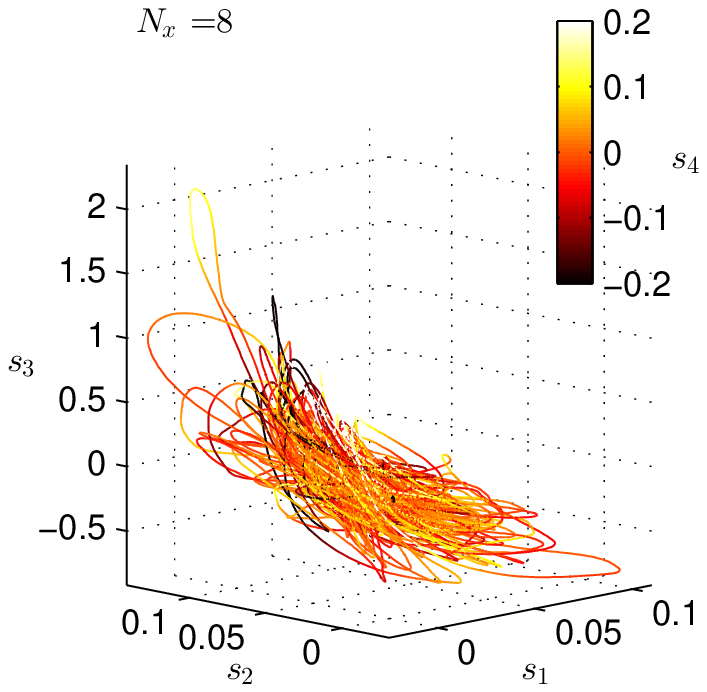}}
\end{center}
\caption{(Color online) Phase-space plots of the 2-D simulations at ($x_0$,$y_0$) with increasing $N_x$ = 1, 2, 4, 8.}
\label{fig.4}
\end{figure*}

For $N_x = 1$, the system settles into a limit cycle pattern with three (relatively) uniformly spaced waves in the domain. Fig.~\ref{fig.3} shows a snapshot of the concentration pattern (far left) and the nearly identical waves passing a probe point at ($x_0$,$y_0$) = ($d^*_x/2$, $L^*_y/2$), which appear incredibly qualitatively similar to the fixed-flux TFM bubbly void waves (e.g., see Fig.~\ref{fig.4} of Ref. \cite{robinson08}). In order to visualize the limit cycle, consider the following four state variables: 
\beqa
{s_1} &=& \phi  - \left\langle \phi  \right\rangle  \nn
{s_2} &=& \frac{{{\rho _g}{d_p}}}{{{\mu _g}}}\sqrt T  - Re_T^{({\mathcal H})} \nn
{s_3} &=& \frac{{{\rho _g}{d_p}}}{{{\mu _g}}}(1 - \phi )({v_g} - {v_s}) - Re_m^{({\mathcal H})} \nn
{s_4} &=& \frac{{{\rho _g}{d_p}}}{{{\mu _g}}}(1 - \phi )({u_g} - {u_s}) \nonumber
\label{eq.statevars}
\eeqa
with state variables evaluated at ($x_0$,$y_0$). The ${\mathcal H}$ superscripts indicate the homogeneous solution \cite{fullmer17} and $Re_T$ and $Re_m$ are the thermal- and mean-flow Reynolds numbers \cite{garzo12, fullmer17}. The linearly stable solution coincides with $s_1 = s_2 = s_3 = s_4 = 0$, hence these four state variables provide a local measure of inhomogeneity. The $s_1-s_2-s_3$ phase-space colored by $s_4$ is shown in Fig.~\ref{fig.4}; the black sphere denotes the origin. For the case $N_x = 1$, the data at ($x_0$,$y_0$) collapse on a limit cycle with no transverse features.

With just two transverse nodes, the one-dimensional limit cycle behavior is destroyed as illustrated in Fig.~\ref{fig.4}, although this case would need further study to say whether or not this case is chaotic or toroidal. However, the instantaneous wave pattern for $N_x = 2$ displayed in Fig.~\ref{fig.3} already hints at the distinction between periodic and chaotic behavior: i) the waves are not uniformly spaced, ii) the waves are not of equal amplitude and, most importantly iii) the lower of the three waves is not uniform in the transverse dimension. This behavior is more easily observed in Fig.~\ref{fig.3} with $N_x$ = 4 and 8 nodes. In the (relatively) wider systems, transverse with spanning plugs also form, at least temporarily, analogous to the 1-D predictions. However, a secondary instability \cite{batchelor93, anderson95, glasser97, glasser98, duru02, sundaresan03} is then triggered which causes the plug to have a multi-dimensional funnel shaped structure which can be seen in Fig.~\ref{fig.3}. Ultimately, the plug spills through the funnel or, equivalently (depending on the readers preferred frame reference), the downstream bubble pushes through and breaks up the plug. The process can be qualitatively perceived as a translating Rayleigh-Taylor instability. In the phase-space of Fig.~\ref{fig.4}, the periodic behavior is destroyed entirely, the origin becomes obscured by the entangled dynamics and their magnitude grows away from the origin.

Although the root cause of clustering is a kinematic instability, that alone appears to be insufficient to cause chaotic behavior. Therefore, when the secondary instability is eliminated or suppressed through dimensional or domain-size reductions, only limit cycle behavior may be expected – at least for continuum predictions of multi-phase flow instabilities. It should further be noted that, while necessary, a multidimensional domain is not a sufficient condition for chaos. The governing conditions of the system studied here were sufficiently unstable for the multidimensional domain to be chaotic; other conditions, notably at higher $Ar$, may only yield periodic behavior even in multidimensional domains \cite{gevrin10, fullmerJFM}.

It is worth noting that the findings in this work carry some practical consequences for KT-TFM users. For one, chaotic behavior means that one cannot simply look at a single instance in time for quantitative information. Simulations need to be either averaged over a sufficiently long period of time or run as several realizations and ensemble averaged. This consequence is not very surprising, because, whether the dynamics were periodic or chaotic, it is universally known that most solutions are temporal and sufficiently complex to require averaging of measures and statistics. The second practical consequence is that slight differences between simulations -- no matter how small, so long as the difference is representable by the precision of the floating point operations of the numerical solution -- will eventually diverge into different instantaneous solutions. Differences may be something rather obvious like different initial conditions, different grids, different time steps, different numerical schemes, different tolerance limits, etc. However, solution differences may also result from unexpected sources such as compiling the same code on different machines, using different compilers or potentially something as simple as different compiler options. In closing, we remark that since bubbles in a dense emulsion seem to stem from the same instability mechanisms as clusters in a dilute suspension \cite{glasser98}, continuum predictions of bubbling gas-solid fluidized beds are also likely chaotic -- although such a model would need to be augmented with several of the closures specifically neglected in this study.

\acknowledgments
The authors are foremost grateful to Dr. Sofiane Benyahia; this study would have never been undertaken without his honest and forthright discussions with the MFiX community. We are grateful for the funding support provided by the U.S. Department of Energy under Grant No. DE-FE0026298.

%\appendix
%\section{Numerical Method}
%\label{appx.num}
%The appendix material goes here.  

%\bibliographystyle{apsrev}
\bibliography{all}

%merlin.mbs aipnum4-1.bst 2010-07-25 4.21a (PWD, AO, DPC) hacked
%Control: key (0)
%Control: author (8) initials jnrlst
%Control: editor formatted (1) identically to author
%Control: production of article title (0) allowed
%Control: page (1) range
%Control: year (1) truncated
%Control: production of eprint (0) enabled
\begin{thebibliography}{55}%
\makeatletter
\providecommand \@ifxundefined [1]{%
 \@ifx{#1\undefined}
}%
\providecommand \@ifnum [1]{%
 \ifnum #1\expandafter \@firstoftwo
 \else \expandafter \@secondoftwo
 \fi
}%
\providecommand \@ifx [1]{%
 \ifx #1\expandafter \@firstoftwo
 \else \expandafter \@secondoftwo
 \fi
}%
\providecommand \natexlab [1]{#1}%
\providecommand \enquote  [1]{``#1''}%
\providecommand \bibnamefont  [1]{#1}%
\providecommand \bibfnamefont [1]{#1}%
\providecommand \citenamefont [1]{#1}%
\providecommand \href@noop [0]{\@secondoftwo}%
\providecommand \href [0]{\begingroup \@sanitize@url \@href}%
\providecommand \@href[1]{\@@startlink{#1}\@@href}%
\providecommand \@@href[1]{\endgroup#1\@@endlink}%
\providecommand \@sanitize@url [0]{\catcode `\\12\catcode `\$12\catcode
  `\&12\catcode `\#12\catcode `\^12\catcode `\_12\catcode `\%12\relax}%
\providecommand \@@startlink[1]{}%
\providecommand \@@endlink[0]{}%
\providecommand \url  [0]{\begingroup\@sanitize@url \@url }%
\providecommand \@url [1]{\endgroup\@href {#1}{\urlprefix }}%
\providecommand \urlprefix  [0]{URL }%
\providecommand \Eprint [0]{\href }%
\providecommand \doibase [0]{http://dx.doi.org/}%
\providecommand \selectlanguage [0]{\@gobble}%
\providecommand \bibinfo  [0]{\@secondoftwo}%
\providecommand \bibfield  [0]{\@secondoftwo}%
\providecommand \translation [1]{[#1]}%
\providecommand \BibitemOpen [0]{}%
\providecommand \bibitemStop [0]{}%
\providecommand \bibitemNoStop [0]{.\EOS\space}%
\providecommand \EOS [0]{\spacefactor3000\relax}%
\providecommand \BibitemShut  [1]{\csname bibitem#1\endcsname}%
\let\auto@bib@innerbib\@empty
%</preamble>
\bibitem [{\citenamefont {Jaeger}, \citenamefont {Nagel},\ and\ \citenamefont
  {Behringer}(1996)}]{jaeger96}%
  \BibitemOpen
  \bibfield  {author} {\bibinfo {author} {\bibfnamefont {H.~M.}\ \bibnamefont
  {Jaeger}}, \bibinfo {author} {\bibfnamefont {S.~R.}\ \bibnamefont {Nagel}}, \
  and\ \bibinfo {author} {\bibfnamefont {R.~P.}\ \bibnamefont {Behringer}},\
  }\bibfield  {title} {\enquote {\bibinfo {title} {Granular solids, liquids,
  and gases},}\ }\href@noop {} {\bibfield  {journal} {\bibinfo  {journal}
  {Reviews of Modern Physics}\ }\textbf {\bibinfo {volume} {68}},\ \bibinfo
  {pages} {1259--1273} (\bibinfo {year} {1996})}\BibitemShut {NoStop}%
\bibitem [{\citenamefont {Geng}\ \emph {et~al.}(2001)\citenamefont {Geng},
  \citenamefont {Howell}, \citenamefont {Longhi}, \citenamefont {Behringer},
  \citenamefont {Reydellet}, \citenamefont {Vanel}, \citenamefont {Clement},\
  and\ \citenamefont {Luding}}]{geng01}%
  \BibitemOpen
  \bibfield  {author} {\bibinfo {author} {\bibfnamefont {J.~F.}\ \bibnamefont
  {Geng}}, \bibinfo {author} {\bibfnamefont {D.}~\bibnamefont {Howell}},
  \bibinfo {author} {\bibfnamefont {E.}~\bibnamefont {Longhi}}, \bibinfo
  {author} {\bibfnamefont {R.~P.}\ \bibnamefont {Behringer}}, \bibinfo {author}
  {\bibfnamefont {G.}~\bibnamefont {Reydellet}}, \bibinfo {author}
  {\bibfnamefont {L.}~\bibnamefont {Vanel}}, \bibinfo {author} {\bibfnamefont
  {E.}~\bibnamefont {Clement}}, \ and\ \bibinfo {author} {\bibfnamefont
  {S.}~\bibnamefont {Luding}},\ }\bibfield  {title} {\enquote {\bibinfo {title}
  {Footprints in sand: The response of a granular material to local
  perturbations},}\ }\href@noop {} {\bibfield  {journal} {\bibinfo  {journal}
  {Physical review letters}\ }\textbf {\bibinfo {volume} {87}} (\bibinfo {year}
  {2001})}\BibitemShut {NoStop}%
\bibitem [{\citenamefont {Forterre}\ and\ \citenamefont
  {Pouliquen}(2008)}]{forterre08}%
  \BibitemOpen
  \bibfield  {author} {\bibinfo {author} {\bibfnamefont {Y.}~\bibnamefont
  {Forterre}}\ and\ \bibinfo {author} {\bibfnamefont {O.}~\bibnamefont
  {Pouliquen}},\ }\bibfield  {title} {\enquote {\bibinfo {title} {Flows of
  dense granular media},}\ }\href@noop {} {\bibfield  {journal} {\bibinfo
  {journal} {Annual Review of Fluid Mechanics}\ }\textbf {\bibinfo {volume}
  {40}},\ \bibinfo {pages} {1--24} (\bibinfo {year} {2008})}\BibitemShut
  {NoStop}%
\bibitem [{\citenamefont {Goldhirsch}(2003)}]{goldhirsch03}%
  \BibitemOpen
  \bibfield  {author} {\bibinfo {author} {\bibfnamefont {I.}~\bibnamefont
  {Goldhirsch}},\ }\bibfield  {title} {\enquote {\bibinfo {title} {Rapid
  granular flows},}\ }\href@noop {} {\bibfield  {journal} {\bibinfo  {journal}
  {Ann. Rev. Fluid Mech.}\ }\textbf {\bibinfo {volume} {35}},\ \bibinfo {pages}
  {267--293} (\bibinfo {year} {2003})}\BibitemShut {NoStop}%
\bibitem [{\citenamefont {Kunii}\ and\ \citenamefont
  {Levenspiel}(1991)}]{kunii}%
  \BibitemOpen
  \bibfield  {author} {\bibinfo {author} {\bibfnamefont {D.~o.}\ \bibnamefont
  {Kunii}}\ and\ \bibinfo {author} {\bibfnamefont {O.}~\bibnamefont
  {Levenspiel}},\ }\href@noop {} {\emph {\bibinfo {title} {Fluidization
  engineering}}},\ \bibinfo {edition} {2nd}\ ed.,\ Butterworth-Heinemann series
  in chemical engineering\ (\bibinfo  {publisher} {Butterworth-Heinemann},\
  \bibinfo {address} {Boston},\ \bibinfo {year} {1991})\BibitemShut {NoStop}%
\bibitem [{\citenamefont {Fullmer}\ and\ \citenamefont
  {Hrenya}(2017)}]{fullmer17}%
  \BibitemOpen
  \bibfield  {author} {\bibinfo {author} {\bibfnamefont {W.~D.}\ \bibnamefont
  {Fullmer}}\ and\ \bibinfo {author} {\bibfnamefont {C.~M.}\ \bibnamefont
  {Hrenya}},\ }\bibfield  {title} {\enquote {\bibinfo {title} {The clustering
  instability in rapid granular and gas-solid flows},}\ }\href@noop {}
  {\bibfield  {journal} {\bibinfo  {journal} {Annual Review of Fluid
  Mechanics}\ }\textbf {\bibinfo {volume} {49}},\ \bibinfo {pages} {485--510}
  (\bibinfo {year} {2017})}\BibitemShut {NoStop}%
\bibitem [{\citenamefont {Agrawal}\ \emph {et~al.}(2001)\citenamefont
  {Agrawal}, \citenamefont {Loezos}, \citenamefont {Syamlal},\ and\
  \citenamefont {Sundaresan}}]{agrawal01}%
  \BibitemOpen
  \bibfield  {author} {\bibinfo {author} {\bibfnamefont {K.}~\bibnamefont
  {Agrawal}}, \bibinfo {author} {\bibfnamefont {P.~N.}\ \bibnamefont {Loezos}},
  \bibinfo {author} {\bibfnamefont {M.}~\bibnamefont {Syamlal}}, \ and\
  \bibinfo {author} {\bibfnamefont {S.}~\bibnamefont {Sundaresan}},\ }\bibfield
   {title} {\enquote {\bibinfo {title} {The role of meso-scale structures in
  rapid gas–solid flows},}\ }\href@noop {} {\bibfield  {journal} {\bibinfo
  {journal} {Journal of Fluid Mechanics}\ }\textbf {\bibinfo {volume} {445}},\
  \bibinfo {pages} {151--185} (\bibinfo {year} {2001})}\BibitemShut {NoStop}%
\bibitem [{\citenamefont {Mitrano}\ \emph {et~al.}(2014)\citenamefont
  {Mitrano}, \citenamefont {Zenk}, \citenamefont {Benyahia}, \citenamefont
  {Galvin}, \citenamefont {Dahl},\ and\ \citenamefont {Hrenya}}]{mitrano14}%
  \BibitemOpen
  \bibfield  {author} {\bibinfo {author} {\bibfnamefont {P.~P.}\ \bibnamefont
  {Mitrano}}, \bibinfo {author} {\bibfnamefont {J.~R.}\ \bibnamefont {Zenk}},
  \bibinfo {author} {\bibfnamefont {S.}~\bibnamefont {Benyahia}}, \bibinfo
  {author} {\bibfnamefont {J.~E.}\ \bibnamefont {Galvin}}, \bibinfo {author}
  {\bibfnamefont {S.~R.}\ \bibnamefont {Dahl}}, \ and\ \bibinfo {author}
  {\bibfnamefont {C.~M.}\ \bibnamefont {Hrenya}},\ }\bibfield  {title}
  {\enquote {\bibinfo {title} {Kinetic-theory predictions of clustering
  instabilities in granular flows: beyond the small-knudsen-number regime},}\
  }\href@noop {} {\bibfield  {journal} {\bibinfo  {journal} {Journal of Fluid
  Mechanics}\ }\textbf {\bibinfo {volume} {738}},\ \bibinfo {pages} {R2--12}
  (\bibinfo {year} {2014})}\BibitemShut {NoStop}%
\bibitem [{\citenamefont {Fullmer}\ and\ \citenamefont
  {Hrenya}(2016)}]{fullmer16}%
  \BibitemOpen
  \bibfield  {author} {\bibinfo {author} {\bibfnamefont {W.~D.}\ \bibnamefont
  {Fullmer}}\ and\ \bibinfo {author} {\bibfnamefont {C.~M.}\ \bibnamefont
  {Hrenya}},\ }\bibfield  {title} {\enquote {\bibinfo {title} {Quantitative
  assessment of fine-grid kinetic-theory-based predictions of mean-slip in
  unbounded fluidization},}\ }\href@noop {} {\bibfield  {journal} {\bibinfo
  {journal} {AIChE Journal}\ }\textbf {\bibinfo {volume} {62}},\ \bibinfo
  {pages} {11--17} (\bibinfo {year} {2016})}\BibitemShut {NoStop}%
\bibitem [{\citenamefont {Fullmer}\ \emph {et~al.}(tted)\citenamefont
  {Fullmer}, \citenamefont {Liu}, \citenamefont {Yin},\ and\ \citenamefont
  {Hrenya}}]{fullmerJFM}%
  \BibitemOpen
  \bibfield  {author} {\bibinfo {author} {\bibfnamefont {W.~D.}\ \bibnamefont
  {Fullmer}}, \bibinfo {author} {\bibfnamefont {G.}~\bibnamefont {Liu}},
  \bibinfo {author} {\bibfnamefont {X.}~\bibnamefont {Yin}}, \ and\ \bibinfo
  {author} {\bibfnamefont {C.~M.}\ \bibnamefont {Hrenya}},\ }\bibfield  {title}
  {\enquote {\bibinfo {title} {Clustering instabilities in sedimenting
  fluid-solid systems: Critical assessment of kinetic-theory-based predictions
  using dns data},}\ }\href@noop {} {\bibfield  {journal} {\bibinfo  {journal}
  {Journal of Fluid Mechanics}\ } (\bibinfo {year} {submitted})}\BibitemShut
  {NoStop}%
\bibitem [{\citenamefont {Hopkins}\ and\ \citenamefont
  {Louge}(1991)}]{hopkins91}%
  \BibitemOpen
  \bibfield  {author} {\bibinfo {author} {\bibfnamefont {M.~A.}\ \bibnamefont
  {Hopkins}}\ and\ \bibinfo {author} {\bibfnamefont {M.~Y.}\ \bibnamefont
  {Louge}},\ }\bibfield  {title} {\enquote {\bibinfo {title} {Inelastic
  microstructure in rapid granular flows of smooth disks},}\ }\href@noop {}
  {\bibfield  {journal} {\bibinfo  {journal} {Physics of Fluids A - Fluid
  Dynamics}\ }\textbf {\bibinfo {volume} {3}},\ \bibinfo {pages} {47--57}
  (\bibinfo {year} {1991})}\BibitemShut {NoStop}%
\bibitem [{\citenamefont {Goldhirsch}\ and\ \citenamefont
  {Zanetti}(1993)}]{goldhirsch93}%
  \BibitemOpen
  \bibfield  {author} {\bibinfo {author} {\bibfnamefont {I.}~\bibnamefont
  {Goldhirsch}}\ and\ \bibinfo {author} {\bibfnamefont {G.}~\bibnamefont
  {Zanetti}},\ }\bibfield  {title} {\enquote {\bibinfo {title} {Clustering
  instability in dissipative gases},}\ }\href@noop {} {\bibfield  {journal}
  {\bibinfo  {journal} {Physical review letters}\ }\textbf {\bibinfo {volume}
  {70}},\ \bibinfo {pages} {1619--1622} (\bibinfo {year} {1993})}\BibitemShut
  {NoStop}%
\bibitem [{\citenamefont {Goldfarb}, \citenamefont {Glasser},\ and\
  \citenamefont {Shinbrot}(2002)}]{goldfarb02}%
  \BibitemOpen
  \bibfield  {author} {\bibinfo {author} {\bibfnamefont {D.~J.}\ \bibnamefont
  {Goldfarb}}, \bibinfo {author} {\bibfnamefont {B.~J.}\ \bibnamefont
  {Glasser}}, \ and\ \bibinfo {author} {\bibfnamefont {T.}~\bibnamefont
  {Shinbrot}},\ }\bibfield  {title} {\enquote {\bibinfo {title} {Shear
  instabilities in granular flows},}\ }\href@noop {} {\bibfield  {journal}
  {\bibinfo  {journal} {Nature}\ }\textbf {\bibinfo {volume} {415}},\ \bibinfo
  {pages} {302--305} (\bibinfo {year} {2002})}\BibitemShut {NoStop}%
\bibitem [{\citenamefont {Conway}, \citenamefont {Shinbrot},\ and\
  \citenamefont {Glasser}(2004)}]{conway04}%
  \BibitemOpen
  \bibfield  {author} {\bibinfo {author} {\bibfnamefont {S.~L.}\ \bibnamefont
  {Conway}}, \bibinfo {author} {\bibfnamefont {T.}~\bibnamefont {Shinbrot}}, \
  and\ \bibinfo {author} {\bibfnamefont {B.~J.}\ \bibnamefont {Glasser}},\
  }\bibfield  {title} {\enquote {\bibinfo {title} {A taylor vortex analogy in
  granular flows},}\ }\href@noop {} {\bibfield  {journal} {\bibinfo  {journal}
  {Nature}\ }\textbf {\bibinfo {volume} {431}},\ \bibinfo {pages} {433--437}
  (\bibinfo {year} {2004})}\BibitemShut {NoStop}%
\bibitem [{\citenamefont {Aranson}\ and\ \citenamefont
  {Tsimring}(2006)}]{aranson06}%
  \BibitemOpen
  \bibfield  {author} {\bibinfo {author} {\bibfnamefont {I.~S.}\ \bibnamefont
  {Aranson}}\ and\ \bibinfo {author} {\bibfnamefont {L.~S.}\ \bibnamefont
  {Tsimring}},\ }\bibfield  {title} {\enquote {\bibinfo {title} {Patterns and
  collective behavior in granular media: Theoretical concepts},}\ }\href@noop
  {} {\bibfield  {journal} {\bibinfo  {journal} {Reviews of Modern Physics}\
  }\textbf {\bibinfo {volume} {78}},\ \bibinfo {pages} {641--692} (\bibinfo
  {year} {2006})}\BibitemShut {NoStop}%
\bibitem [{\citenamefont {Vinningland}\ \emph {et~al.}(2007)\citenamefont
  {Vinningland}, \citenamefont {Johnsen}, \citenamefont {Flekkoy},
  \citenamefont {Toussaint},\ and\ \citenamefont {Maloy}}]{vinningland07}%
  \BibitemOpen
  \bibfield  {author} {\bibinfo {author} {\bibfnamefont {J.~L.}\ \bibnamefont
  {Vinningland}}, \bibinfo {author} {\bibfnamefont {O.}~\bibnamefont
  {Johnsen}}, \bibinfo {author} {\bibfnamefont {E.~G.}\ \bibnamefont
  {Flekkoy}}, \bibinfo {author} {\bibfnamefont {R.}~\bibnamefont {Toussaint}},
  \ and\ \bibinfo {author} {\bibfnamefont {K.~J.}\ \bibnamefont {Maloy}},\
  }\bibfield  {title} {\enquote {\bibinfo {title} {Granular rayleigh-taylor
  instability: Experiments and simulations},}\ }\href@noop {} {\bibfield
  {journal} {\bibinfo  {journal} {Physical review letters}\ }\textbf {\bibinfo
  {volume} {99}} (\bibinfo {year} {2007})}\BibitemShut {NoStop}%
\bibitem [{\citenamefont {Cheng}\ \emph {et~al.}(2008)\citenamefont {Cheng},
  \citenamefont {Xu}, \citenamefont {Patterson}, \citenamefont {Jaeger},\ and\
  \citenamefont {Nagel}}]{cheng08}%
  \BibitemOpen
  \bibfield  {author} {\bibinfo {author} {\bibfnamefont {X.}~\bibnamefont
  {Cheng}}, \bibinfo {author} {\bibfnamefont {L.}~\bibnamefont {Xu}}, \bibinfo
  {author} {\bibfnamefont {A.}~\bibnamefont {Patterson}}, \bibinfo {author}
  {\bibfnamefont {H.~M.}\ \bibnamefont {Jaeger}}, \ and\ \bibinfo {author}
  {\bibfnamefont {S.~R.}\ \bibnamefont {Nagel}},\ }\bibfield  {title} {\enquote
  {\bibinfo {title} {Towards the zero-surface-tension limit in granular
  fingering instability},}\ }\href@noop {} {\bibfield  {journal} {\bibinfo
  {journal} {Nature Physics}\ }\textbf {\bibinfo {volume} {4}},\ \bibinfo
  {pages} {234--237} (\bibinfo {year} {2008})}\BibitemShut {NoStop}%
\bibitem [{\citenamefont {Christov}, \citenamefont {Ottino},\ and\
  \citenamefont {Lueptow}(2010)}]{christov10}%
  \BibitemOpen
  \bibfield  {author} {\bibinfo {author} {\bibfnamefont {I.~C.}\ \bibnamefont
  {Christov}}, \bibinfo {author} {\bibfnamefont {J.~M.}\ \bibnamefont
  {Ottino}}, \ and\ \bibinfo {author} {\bibfnamefont {R.~M.}\ \bibnamefont
  {Lueptow}},\ }\bibfield  {title} {\enquote {\bibinfo {title} {Chaotic mixing
  via streamline jumping in quasi-two-dimensional tumbled granular flows},}\
  }\href@noop {} {\bibfield  {journal} {\bibinfo  {journal} {Chaos}\ }\textbf
  {\bibinfo {volume} {20}} (\bibinfo {year} {2010})}\BibitemShut {NoStop}%
\bibitem [{\citenamefont {Seiden}\ and\ \citenamefont
  {Thomas}(2011)}]{seiden11}%
  \BibitemOpen
  \bibfield  {author} {\bibinfo {author} {\bibfnamefont {G.}~\bibnamefont
  {Seiden}}\ and\ \bibinfo {author} {\bibfnamefont {P.~J.}\ \bibnamefont
  {Thomas}},\ }\bibfield  {title} {\enquote {\bibinfo {title} {Complexity,
  segregation, and pattern formation in rotating-drum flows},}\ }\href@noop {}
  {\bibfield  {journal} {\bibinfo  {journal} {Reviews of Modern Physics}\
  }\textbf {\bibinfo {volume} {83}},\ \bibinfo {pages} {1323--1365} (\bibinfo
  {year} {2011})}\BibitemShut {NoStop}%
\bibitem [{\citenamefont {Shinbrot}(2015)}]{shinbrot15}%
  \BibitemOpen
  \bibfield  {author} {\bibinfo {author} {\bibfnamefont {T.}~\bibnamefont
  {Shinbrot}},\ }\bibfield  {title} {\enquote {\bibinfo {title} {Granular chaos
  and mixing: Whirled in a grain of sand},}\ }\href@noop {} {\bibfield
  {journal} {\bibinfo  {journal} {Chaos}\ }\textbf {\bibinfo {volume} {25}}
  (\bibinfo {year} {2015})}\BibitemShut {NoStop}%
\bibitem [{\citenamefont {Cocco}, \citenamefont {Karri},\ and\ \citenamefont
  {Knowlton}(2014)}]{cocco14}%
  \BibitemOpen
  \bibfield  {author} {\bibinfo {author} {\bibfnamefont {R.}~\bibnamefont
  {Cocco}}, \bibinfo {author} {\bibfnamefont {S.~B.~R.}\ \bibnamefont {Karri}},
  \ and\ \bibinfo {author} {\bibfnamefont {T.}~\bibnamefont {Knowlton}},\
  }\bibfield  {title} {\enquote {\bibinfo {title} {Introduction to
  fluidization},}\ }\href@noop {} {\bibfield  {journal} {\bibinfo  {journal}
  {Chemical Engineering Progress}\ }\textbf {\bibinfo {volume} {110}},\
  \bibinfo {pages} {21--29} (\bibinfo {year} {2014})}\BibitemShut {NoStop}%
\bibitem [{\citenamefont {van~der Hoef}\ \emph {et~al.}(2008)\citenamefont
  {van~der Hoef}, \citenamefont {van Sint~Annaland}, \citenamefont {Deen},\
  and\ \citenamefont {Kuipers}}]{vanderhoef08}%
  \BibitemOpen
  \bibfield  {author} {\bibinfo {author} {\bibfnamefont {M.}~\bibnamefont
  {van~der Hoef}}, \bibinfo {author} {\bibfnamefont {M.}~\bibnamefont {van
  Sint~Annaland}}, \bibinfo {author} {\bibfnamefont {N.}~\bibnamefont {Deen}},
  \ and\ \bibinfo {author} {\bibfnamefont {J.}~\bibnamefont {Kuipers}},\
  }\bibfield  {title} {\enquote {\bibinfo {title} {Numerical simulation of
  dense gas-solid fluidized beds: A multiscale modeling strategy},}\
  }\href@noop {} {\bibfield  {journal} {\bibinfo  {journal} {Annu. Rev. Fluid
  Mech.}\ }\textbf {\bibinfo {volume} {40}},\ \bibinfo {pages} {47--70}
  (\bibinfo {year} {2008})}\BibitemShut {NoStop}%
\bibitem [{\citenamefont {Li}\ \emph {et~al.}(2013)\citenamefont {Li},
  \citenamefont {Ge}, \citenamefont {Wang}, \citenamefont {Yang}, \citenamefont
  {Liu}, \citenamefont {Wang}, \citenamefont {He}, \citenamefont {Wang},
  \citenamefont {Wang},\ and\ \citenamefont {Guo}}]{li13}%
  \BibitemOpen
  \bibfield  {author} {\bibinfo {author} {\bibfnamefont {J.}~\bibnamefont
  {Li}}, \bibinfo {author} {\bibfnamefont {W.}~\bibnamefont {Ge}}, \bibinfo
  {author} {\bibfnamefont {W.}~\bibnamefont {Wang}}, \bibinfo {author}
  {\bibfnamefont {N.}~\bibnamefont {Yang}}, \bibinfo {author} {\bibfnamefont
  {X.}~\bibnamefont {Liu}}, \bibinfo {author} {\bibfnamefont {L.}~\bibnamefont
  {Wang}}, \bibinfo {author} {\bibfnamefont {X.}~\bibnamefont {He}}, \bibinfo
  {author} {\bibfnamefont {X.}~\bibnamefont {Wang}}, \bibinfo {author}
  {\bibfnamefont {J.}~\bibnamefont {Wang}}, \ and\ \bibinfo {author}
  {\bibfnamefont {M.}~\bibnamefont {Guo}},\ }\href@noop {} {\emph {\bibinfo
  {title} {From multiscale modeling to meso-science : a chemical engineering
  perspective : principles, modeling, simulation, and application}}}\ (\bibinfo
   {publisher} {Springer-Verlag},\ \bibinfo {address} {Berlin; Heidelberg},\
  \bibinfo {year} {2013})\BibitemShut {NoStop}%
\bibitem [{\citenamefont {Cocco}\ \emph {et~al.}(2017)\citenamefont {Cocco},
  \citenamefont {Fullmer}, \citenamefont {Liu},\ and\ \citenamefont
  {Hrenya}}]{cocco17}%
  \BibitemOpen
  \bibfield  {author} {\bibinfo {author} {\bibfnamefont {R.}~\bibnamefont
  {Cocco}}, \bibinfo {author} {\bibfnamefont {W.~D.}\ \bibnamefont {Fullmer}},
  \bibinfo {author} {\bibfnamefont {P.}~\bibnamefont {Liu}}, \ and\ \bibinfo
  {author} {\bibfnamefont {C.~M.}\ \bibnamefont {Hrenya}},\ }\bibfield  {title}
  {\enquote {\bibinfo {title} {{CFD-DEM}: Modeling the small to understand the
  large},}\ }\href@noop {} {\bibfield  {journal} {\bibinfo  {journal} {Chemical
  Engineering Progress}\ ,\ \bibinfo {pages} {in press}} (\bibinfo {year}
  {2017})}\BibitemShut {NoStop}%
\bibitem [{\citenamefont {Gidaspow}(1994)}]{gidaspow}%
  \BibitemOpen
  \bibfield  {author} {\bibinfo {author} {\bibfnamefont {D.}~\bibnamefont
  {Gidaspow}},\ }\href@noop {} {\emph {\bibinfo {title} {Multiphase Flow and
  Fluidization}}}\ (\bibinfo  {publisher} {Academic Press},\ \bibinfo {address}
  {San Diego},\ \bibinfo {year} {1994})\BibitemShut {NoStop}%
\bibitem [{\citenamefont {Brilliantov}\ and\ \citenamefont
  {P\"{o}schel}(2004)}]{brilliantov}%
  \BibitemOpen
  \bibfield  {author} {\bibinfo {author} {\bibfnamefont {N.}~\bibnamefont
  {Brilliantov}}\ and\ \bibinfo {author} {\bibfnamefont {T.}~\bibnamefont
  {P\"{o}schel}},\ }\href@noop {} {\emph {\bibinfo {title} {Kinetic theory of
  granular gases}}}\ (\bibinfo  {publisher} {Oxford University Press},\
  \bibinfo {address} {Oxford ; New York},\ \bibinfo {year} {2004})\BibitemShut
  {NoStop}%
\bibitem [{\citenamefont {Pannala}, \citenamefont {Syamlal},\ and\
  \citenamefont {O'Brien}(2010)}]{pannala}%
  \BibitemOpen
  \bibfield  {author} {\bibinfo {author} {\bibfnamefont {S.}~\bibnamefont
  {Pannala}}, \bibinfo {author} {\bibfnamefont {M.}~\bibnamefont {Syamlal}}, \
  and\ \bibinfo {author} {\bibfnamefont {T.}~\bibnamefont {O'Brien}},\
  }\href@noop {} {\emph {\bibinfo {title} {Computational gas-solid flows and
  reacting systems: Theory, Methods and Practice}}}\ (\bibinfo  {publisher}
  {IGI Global},\ \bibinfo {address} {Hershey},\ \bibinfo {year}
  {2010})\BibitemShut {NoStop}%
\bibitem [{\citenamefont {Yin}\ \emph {et~al.}(2013)\citenamefont {Yin},
  \citenamefont {Zenk}, \citenamefont {Mitrano},\ and\ \citenamefont
  {Hrenya}}]{yin13}%
  \BibitemOpen
  \bibfield  {author} {\bibinfo {author} {\bibfnamefont {X.}~\bibnamefont
  {Yin}}, \bibinfo {author} {\bibfnamefont {J.~R.}\ \bibnamefont {Zenk}},
  \bibinfo {author} {\bibfnamefont {P.~P.}\ \bibnamefont {Mitrano}}, \ and\
  \bibinfo {author} {\bibfnamefont {C.~M.}\ \bibnamefont {Hrenya}},\ }\bibfield
   {title} {\enquote {\bibinfo {title} {Impact of collisional versus viscous
  dissipation on flow instabilities in gas–solid systems},}\ }\href@noop {}
  {\bibfield  {journal} {\bibinfo  {journal} {Journal of Fluid Mechanics}\
  }\textbf {\bibinfo {volume} {727}},\ \bibinfo {pages} {R2} (\bibinfo {year}
  {2013})}\BibitemShut {NoStop}%
\bibitem [{\citenamefont {Radl}\ and\ \citenamefont
  {Sundaresan}(2014)}]{radl14}%
  \BibitemOpen
  \bibfield  {author} {\bibinfo {author} {\bibfnamefont {S.}~\bibnamefont
  {Radl}}\ and\ \bibinfo {author} {\bibfnamefont {S.}~\bibnamefont
  {Sundaresan}},\ }\bibfield  {title} {\enquote {\bibinfo {title} {A drag model
  for filtered euler-lagrange simulations of clustered gas-particle
  suspensions},}\ }\href@noop {} {\bibfield  {journal} {\bibinfo  {journal}
  {Chemical Engineering Science}\ }\textbf {\bibinfo {volume} {117}},\ \bibinfo
  {pages} {416--425} (\bibinfo {year} {2014})}\BibitemShut {NoStop}%
\bibitem [{\citenamefont {Capecelatro}, \citenamefont {Desjardins},\ and\
  \citenamefont {Fox}(2015)}]{capecelatro15}%
  \BibitemOpen
  \bibfield  {author} {\bibinfo {author} {\bibfnamefont {J.}~\bibnamefont
  {Capecelatro}}, \bibinfo {author} {\bibfnamefont {O.}~\bibnamefont
  {Desjardins}}, \ and\ \bibinfo {author} {\bibfnamefont {R.~O.}\ \bibnamefont
  {Fox}},\ }\bibfield  {title} {\enquote {\bibinfo {title} {On fluid-particle
  dynamics in fully developed cluster-induced turbulence},}\ }\href@noop {}
  {\bibfield  {journal} {\bibinfo  {journal} {Journal of Fluid Mechanics}\
  }\textbf {\bibinfo {volume} {780}},\ \bibinfo {pages} {578--635} (\bibinfo
  {year} {2015})}\BibitemShut {NoStop}%
\bibitem [{\citenamefont {Dellago}\ and\ \citenamefont
  {Posch}(1997)}]{dellago97}%
  \BibitemOpen
  \bibfield  {author} {\bibinfo {author} {\bibfnamefont {C.}~\bibnamefont
  {Dellago}}\ and\ \bibinfo {author} {\bibfnamefont {H.}~\bibnamefont
  {Posch}},\ }\bibfield  {title} {\enquote {\bibinfo {title}
  {Kolmogorov-{S}inai entropy and {L}yapunov spectra of a hard-sphere gas},}\
  }\href@noop {} {\bibfield  {journal} {\bibinfo  {journal} {Physica A}\
  }\textbf {\bibinfo {volume} {240}},\ \bibinfo {pages} {68--83} (\bibinfo
  {year} {1997})}\BibitemShut {NoStop}%
\bibitem [{\citenamefont {McNamara}\ and\ \citenamefont
  {Mareschal}(2001)}]{mcnamara01}%
  \BibitemOpen
  \bibfield  {author} {\bibinfo {author} {\bibfnamefont {S.}~\bibnamefont
  {McNamara}}\ and\ \bibinfo {author} {\bibfnamefont {M.}~\bibnamefont
  {Mareschal}},\ }\bibfield  {title} {\enquote {\bibinfo {title} {Lyapunov
  spectrum of granular gases},}\ }\href@noop {} {\bibfield  {journal} {\bibinfo
   {journal} {Physical Review E}\ }\textbf {\bibinfo {volume} {63}},\ \bibinfo
  {pages} {061306} (\bibinfo {year} {2001})}\BibitemShut {NoStop}%
\bibitem [{\citenamefont {Gevrin}, \citenamefont {Masbernat},\ and\
  \citenamefont {Simonin}(2010)}]{gevrin10}%
  \BibitemOpen
  \bibfield  {author} {\bibinfo {author} {\bibfnamefont {F.}~\bibnamefont
  {Gevrin}}, \bibinfo {author} {\bibfnamefont {O.}~\bibnamefont {Masbernat}}, \
  and\ \bibinfo {author} {\bibfnamefont {O.}~\bibnamefont {Simonin}},\
  }\bibfield  {title} {\enquote {\bibinfo {title} {Numerical study of
  solid-liquid fluidization dynamics},}\ }\href@noop {} {\bibfield  {journal}
  {\bibinfo  {journal} {AIChE Journal}\ }\textbf {\bibinfo {volume} {56}},\
  \bibinfo {pages} {2781--2794} (\bibinfo {year} {2010})}\BibitemShut {NoStop}%
\bibitem [{\citenamefont {Marzocchella}\ \emph {et~al.}(1997)\citenamefont
  {Marzocchella}, \citenamefont {Zijerveld}, \citenamefont {Schouten},\ and\
  \citenamefont {vandenBleek}}]{marzocchella97}%
  \BibitemOpen
  \bibfield  {author} {\bibinfo {author} {\bibfnamefont {A.}~\bibnamefont
  {Marzocchella}}, \bibinfo {author} {\bibfnamefont {R.~C.}\ \bibnamefont
  {Zijerveld}}, \bibinfo {author} {\bibfnamefont {J.~C.}\ \bibnamefont
  {Schouten}}, \ and\ \bibinfo {author} {\bibfnamefont {C.~M.}\ \bibnamefont
  {vandenBleek}},\ }\bibfield  {title} {\enquote {\bibinfo {title} {Chaotic
  behavior of gas-solids flow in the riser of a laboratory-scale circulating
  fluidized bed},}\ }\href@noop {} {\bibfield  {journal} {\bibinfo  {journal}
  {AIChE Journal}\ }\textbf {\bibinfo {volume} {43}},\ \bibinfo {pages}
  {1458--1468} (\bibinfo {year} {1997})}\BibitemShut {NoStop}%
\bibitem [{\citenamefont {Ji}\ \emph {et~al.}(2000)\citenamefont {Ji},
  \citenamefont {Ohara}, \citenamefont {Kuramoto}, \citenamefont {Tsutsumi},
  \citenamefont {Yoshida},\ and\ \citenamefont {Hirama}}]{ji00}%
  \BibitemOpen
  \bibfield  {author} {\bibinfo {author} {\bibfnamefont {H.~S.}\ \bibnamefont
  {Ji}}, \bibinfo {author} {\bibfnamefont {H.}~\bibnamefont {Ohara}}, \bibinfo
  {author} {\bibfnamefont {K.}~\bibnamefont {Kuramoto}}, \bibinfo {author}
  {\bibfnamefont {A.}~\bibnamefont {Tsutsumi}}, \bibinfo {author}
  {\bibfnamefont {K.}~\bibnamefont {Yoshida}}, \ and\ \bibinfo {author}
  {\bibfnamefont {T.}~\bibnamefont {Hirama}},\ }\bibfield  {title} {\enquote
  {\bibinfo {title} {Nonlinear dynamics of gas-solid circulating fluidized-bed
  system},}\ }\href@noop {} {\bibfield  {journal} {\bibinfo  {journal}
  {Chemical Engineering Science}\ }\textbf {\bibinfo {volume} {55}},\ \bibinfo
  {pages} {403--410} (\bibinfo {year} {2000})}\BibitemShut {NoStop}%
\bibitem [{\citenamefont {Benyahia}, \citenamefont {Syamlal},\ and\
  \citenamefont {O'Brien}(2007)}]{benyahia07}%
  \BibitemOpen
  \bibfield  {author} {\bibinfo {author} {\bibfnamefont {S.}~\bibnamefont
  {Benyahia}}, \bibinfo {author} {\bibfnamefont {M.}~\bibnamefont {Syamlal}}, \
  and\ \bibinfo {author} {\bibfnamefont {T.~J.}\ \bibnamefont {O'Brien}},\
  }\bibfield  {title} {\enquote {\bibinfo {title} {Study of the ability of
  multiphase continuum models to predict core-annulus flow},}\ }\href@noop {}
  {\bibfield  {journal} {\bibinfo  {journal} {AIChE J.}\ }\textbf {\bibinfo
  {volume} {53}},\ \bibinfo {pages} {2549--2568} (\bibinfo {year}
  {2007})}\BibitemShut {NoStop}%
\bibitem [{\citenamefont {Benyahia}(2016)}]{benyahia16}%
  \BibitemOpen
  \bibfield  {author} {\bibinfo {author} {\bibfnamefont {S.}~\bibnamefont
  {Benyahia}},\ }\href@noop {} {\enquote {\bibinfo {title} {Personal
  communication},}\ }\bibinfo {howpublished} {E-mail distributed to
  {mfix-help@mfix.netl.doe.gov} on Tue, 26 Jul 2016 18:01:21 +0000} (\bibinfo
  {year} {2016}),\ \bibinfo {note} {archived on sympa for registered users at:
  \url{https://mfix.netl.doe.gov/sympa/arc/mfix-help/2016-07/msg00029.html}}\BibitemShut
  {NoStop}%
\bibitem [{\citenamefont {Robinson}\ \emph {et~al.}(2008)\citenamefont
  {Robinson}, \citenamefont {Fowler}, \citenamefont {Alexander},\ and\
  \citenamefont {O'Brien}}]{robinson08}%
  \BibitemOpen
  \bibfield  {author} {\bibinfo {author} {\bibfnamefont {M.}~\bibnamefont
  {Robinson}}, \bibinfo {author} {\bibfnamefont {A.~C.}\ \bibnamefont
  {Fowler}}, \bibinfo {author} {\bibfnamefont {A.~J.}\ \bibnamefont
  {Alexander}}, \ and\ \bibinfo {author} {\bibfnamefont {S.~B.~G.}\
  \bibnamefont {O'Brien}},\ }\bibfield  {title} {\enquote {\bibinfo {title}
  {Waves in guinness},}\ }\href@noop {} {\bibfield  {journal} {\bibinfo
  {journal} {Physics of Fluids}\ }\textbf {\bibinfo {volume} {20}} (\bibinfo
  {year} {2008})}\BibitemShut {NoStop}%
\bibitem [{\citenamefont {L\'{o}pez~de Bertodano}\ \emph
  {et~al.}(2017)\citenamefont {L\'{o}pez~de Bertodano}, \citenamefont
  {Fullmer}, \citenamefont {Clausse},\ and\ \citenamefont
  {Ransom}}]{bertodano}%
  \BibitemOpen
  \bibfield  {author} {\bibinfo {author} {\bibfnamefont {M.}~\bibnamefont
  {L\'{o}pez~de Bertodano}}, \bibinfo {author} {\bibfnamefont {W.~D.}\
  \bibnamefont {Fullmer}}, \bibinfo {author} {\bibfnamefont {A.}~\bibnamefont
  {Clausse}}, \ and\ \bibinfo {author} {\bibfnamefont {V.~H.}\ \bibnamefont
  {Ransom}},\ }\href@noop {} {\emph {\bibinfo {title} {Two-Fluid Model
  Stability, Simulation and Chaos}}}\ (\bibinfo  {publisher} {Springer
  International Publishing},\ \bibinfo {address} {Switzerland},\ \bibinfo
  {year} {2017})\BibitemShut {NoStop}%
\bibitem [{\citenamefont {Garz\'{o}}\ \emph {et~al.}(2012)\citenamefont
  {Garz\'{o}}, \citenamefont {Tenneti}, \citenamefont {Subramaniam},\ and\
  \citenamefont {Hrenya}}]{garzo12}%
  \BibitemOpen
  \bibfield  {author} {\bibinfo {author} {\bibfnamefont {V.}~\bibnamefont
  {Garz\'{o}}}, \bibinfo {author} {\bibfnamefont {S.}~\bibnamefont {Tenneti}},
  \bibinfo {author} {\bibfnamefont {S.}~\bibnamefont {Subramaniam}}, \ and\
  \bibinfo {author} {\bibfnamefont {C.~M.}\ \bibnamefont {Hrenya}},\ }\bibfield
   {title} {\enquote {\bibinfo {title} {Enskog kinetic theory for monodisperse
  gas–solid flows},}\ }\href@noop {} {\bibfield  {journal} {\bibinfo
  {journal} {Journal of Fluid Mechanics}\ }\textbf {\bibinfo {volume} {712}},\
  \bibinfo {pages} {129--168} (\bibinfo {year} {2012})}\BibitemShut {NoStop}%
\bibitem [{\citenamefont {Beetstra}, \citenamefont {van~der Hoef},\ and\
  \citenamefont {Kuipers}(2007)}]{beetstra07}%
  \BibitemOpen
  \bibfield  {author} {\bibinfo {author} {\bibfnamefont {R.}~\bibnamefont
  {Beetstra}}, \bibinfo {author} {\bibfnamefont {M.~A.}\ \bibnamefont {van~der
  Hoef}}, \ and\ \bibinfo {author} {\bibfnamefont {J.~A.~M.}\ \bibnamefont
  {Kuipers}},\ }\bibfield  {title} {\enquote {\bibinfo {title} {Drag force of
  intermediate reynolds number flow past mono- and bidisperse arrays of
  spheres},}\ }\href@noop {} {\bibfield  {journal} {\bibinfo  {journal} {AIChE
  J.}\ }\textbf {\bibinfo {volume} {53}},\ \bibinfo {pages} {489--501}
  (\bibinfo {year} {2007})}\BibitemShut {NoStop}%
\bibitem [{\citenamefont {Wylie}, \citenamefont {Koch},\ and\ \citenamefont
  {Ladd}(2003)}]{wylie03}%
  \BibitemOpen
  \bibfield  {author} {\bibinfo {author} {\bibfnamefont {J.~J.}\ \bibnamefont
  {Wylie}}, \bibinfo {author} {\bibfnamefont {D.~L.}\ \bibnamefont {Koch}}, \
  and\ \bibinfo {author} {\bibfnamefont {A.~J.~C.}\ \bibnamefont {Ladd}},\
  }\bibfield  {title} {\enquote {\bibinfo {title} {Rheology of suspensions with
  high particle inertia and moderate fluid inertia},}\ }\href@noop {}
  {\bibfield  {journal} {\bibinfo  {journal} {Journal of Fluid Mechanics}\
  }\textbf {\bibinfo {volume} {480}},\ \bibinfo {pages} {95--118} (\bibinfo
  {year} {2003})}\BibitemShut {NoStop}%
\bibitem [{\citenamefont {Carnahan}\ and\ \citenamefont
  {Starling}(1969)}]{carnahan69}%
  \BibitemOpen
  \bibfield  {author} {\bibinfo {author} {\bibfnamefont {N.~F.}\ \bibnamefont
  {Carnahan}}\ and\ \bibinfo {author} {\bibfnamefont {K.~E.}\ \bibnamefont
  {Starling}},\ }\bibfield  {title} {\enquote {\bibinfo {title} {Equation of
  state for nonattracting rigid spheres},}\ }\href@noop {} {\bibfield
  {journal} {\bibinfo  {journal} {Journal of Chemical Physics}\ }\textbf
  {\bibinfo {volume} {51}},\ \bibinfo {pages} {635--636} (\bibinfo {year}
  {1969})}\BibitemShut {NoStop}%
\bibitem [{\citenamefont {Ma}\ and\ \citenamefont {Ahmadi}(1988)}]{ma88}%
  \BibitemOpen
  \bibfield  {author} {\bibinfo {author} {\bibfnamefont {D.}~\bibnamefont
  {Ma}}\ and\ \bibinfo {author} {\bibfnamefont {G.}~\bibnamefont {Ahmadi}},\
  }\bibfield  {title} {\enquote {\bibinfo {title} {A kinetic-model for rapid
  granular flows of nearly elastic particles including interstitial fluid
  effects},}\ }\href@noop {} {\bibfield  {journal} {\bibinfo  {journal} {Powder
  Technology}\ }\textbf {\bibinfo {volume} {56}},\ \bibinfo {pages} {191--207}
  (\bibinfo {year} {1988})}\BibitemShut {NoStop}%
\bibitem [{\citenamefont {Sprott}(2003)}]{sprott}%
  \BibitemOpen
  \bibfield  {author} {\bibinfo {author} {\bibfnamefont {J.~C.}\ \bibnamefont
  {Sprott}},\ }\href@noop {} {\emph {\bibinfo {title} {Chaos and time-series
  analysis}}}\ (\bibinfo  {publisher} {Oxford University Press},\ \bibinfo
  {address} {Oxford ; New York},\ \bibinfo {year} {2003})\BibitemShut {NoStop}%
\bibitem [{\citenamefont {Cencini}, \citenamefont {Cecconi},\ and\
  \citenamefont {Vulpiani}(2010)}]{cencini}%
  \BibitemOpen
  \bibfield  {author} {\bibinfo {author} {\bibfnamefont {M.}~\bibnamefont
  {Cencini}}, \bibinfo {author} {\bibfnamefont {F.}~\bibnamefont {Cecconi}}, \
  and\ \bibinfo {author} {\bibfnamefont {A.}~\bibnamefont {Vulpiani}},\
  }\href@noop {} {\emph {\bibinfo {title} {Chaos : from simple models to
  complex systems}}},\ Series on advances in statistical mechanics\ (\bibinfo
  {publisher} {World Scientific},\ \bibinfo {address} {Hackensack, NJ},\
  \bibinfo {year} {2010})\BibitemShut {NoStop}%
\bibitem [{\citenamefont {Wolf}\ \emph {et~al.}(1985)\citenamefont {Wolf},
  \citenamefont {Swift}, \citenamefont {Swinney},\ and\ \citenamefont
  {Vastano}}]{wolf85}%
  \BibitemOpen
  \bibfield  {author} {\bibinfo {author} {\bibfnamefont {A.}~\bibnamefont
  {Wolf}}, \bibinfo {author} {\bibfnamefont {J.~B.}\ \bibnamefont {Swift}},
  \bibinfo {author} {\bibfnamefont {H.~L.}\ \bibnamefont {Swinney}}, \ and\
  \bibinfo {author} {\bibfnamefont {J.~A.}\ \bibnamefont {Vastano}},\
  }\bibfield  {title} {\enquote {\bibinfo {title} {Determining lyapunov
  exponents from a time-series},}\ }\href@noop {} {\bibfield  {journal}
  {\bibinfo  {journal} {Physica D}\ }\textbf {\bibinfo {volume} {16}},\
  \bibinfo {pages} {285--317} (\bibinfo {year} {1985})}\BibitemShut {NoStop}%
\bibitem [{\citenamefont {Brummitt}\ and\ \citenamefont
  {Sprott}(2009)}]{brummitt09}%
  \BibitemOpen
  \bibfield  {author} {\bibinfo {author} {\bibfnamefont {C.}~\bibnamefont
  {Brummitt}}\ and\ \bibinfo {author} {\bibfnamefont {J.}~\bibnamefont
  {Sprott}},\ }\bibfield  {title} {\enquote {\bibinfo {title} {A search for the
  simplest chaotic partial differential equation},}\ }\href@noop {} {\bibfield
  {journal} {\bibinfo  {journal} {Physics Letters A}\ }\textbf {\bibinfo
  {volume} {373}},\ \bibinfo {pages} {2717--2721} (\bibinfo {year}
  {2009})}\BibitemShut {NoStop}%
\bibitem [{\citenamefont {Fullmer}\ \emph {et~al.}(2014)\citenamefont
  {Fullmer}, \citenamefont {Lopez~de Bertodano}, \citenamefont {Chen},\ and\
  \citenamefont {Clausse}}]{fullmer14b}%
  \BibitemOpen
  \bibfield  {author} {\bibinfo {author} {\bibfnamefont {W.~D.}\ \bibnamefont
  {Fullmer}}, \bibinfo {author} {\bibfnamefont {M.~A.}\ \bibnamefont {Lopez~de
  Bertodano}}, \bibinfo {author} {\bibfnamefont {M.}~\bibnamefont {Chen}}, \
  and\ \bibinfo {author} {\bibfnamefont {A.}~\bibnamefont {Clausse}},\
  }\bibfield  {title} {\enquote {\bibinfo {title} {Analysis of stability,
  verification and chaos with the kreiss--ystr{\"o}m equations},}\ }\href@noop
  {} {\bibfield  {journal} {\bibinfo  {journal} {Applied Mathematics and
  Computation}\ }\textbf {\bibinfo {volume} {248}},\ \bibinfo {pages} {28--46}
  (\bibinfo {year} {2014})}\BibitemShut {NoStop}%
\bibitem [{\citenamefont {Batchelor}(1993)}]{batchelor93}%
  \BibitemOpen
  \bibfield  {author} {\bibinfo {author} {\bibfnamefont {G.~K.}\ \bibnamefont
  {Batchelor}},\ }\bibfield  {title} {\enquote {\bibinfo {title} {Secondary
  instability of a gas-fluidized bed},}\ }\href@noop {} {\bibfield  {journal}
  {\bibinfo  {journal} {Journal of Fluid Mechanics}\ }\textbf {\bibinfo
  {volume} {257}},\ \bibinfo {pages} {359--371} (\bibinfo {year}
  {1993})}\BibitemShut {NoStop}%
\bibitem [{\citenamefont {Anderson}, \citenamefont {Sundaresan},\ and\
  \citenamefont {Jackson}(1995)}]{anderson95}%
  \BibitemOpen
  \bibfield  {author} {\bibinfo {author} {\bibfnamefont {K.}~\bibnamefont
  {Anderson}}, \bibinfo {author} {\bibfnamefont {S.}~\bibnamefont
  {Sundaresan}}, \ and\ \bibinfo {author} {\bibfnamefont {R.}~\bibnamefont
  {Jackson}},\ }\bibfield  {title} {\enquote {\bibinfo {title} {Instabilities
  and the formation of bubbles in fluidized-beds},}\ }\href@noop {} {\bibfield
  {journal} {\bibinfo  {journal} {Journal of Fluid Mechanics}\ }\textbf
  {\bibinfo {volume} {303}},\ \bibinfo {pages} {327--366} (\bibinfo {year}
  {1995})}\BibitemShut {NoStop}%
\bibitem [{\citenamefont {Glasser}, \citenamefont {Kevrekidis},\ and\
  \citenamefont {Sundaresan}(1997)}]{glasser97}%
  \BibitemOpen
  \bibfield  {author} {\bibinfo {author} {\bibfnamefont {B.}~\bibnamefont
  {Glasser}}, \bibinfo {author} {\bibfnamefont {I.}~\bibnamefont {Kevrekidis}},
  \ and\ \bibinfo {author} {\bibfnamefont {S.}~\bibnamefont {Sundaresan}},\
  }\bibfield  {title} {\enquote {\bibinfo {title} {Fully developed travelling
  wave solutions and bubble formation in fluidized beds},}\ }\href@noop {}
  {\bibfield  {journal} {\bibinfo  {journal} {Journal of Fluid Mechanics}\
  }\textbf {\bibinfo {volume} {334}},\ \bibinfo {pages} {157--188} (\bibinfo
  {year} {1997})}\BibitemShut {NoStop}%
\bibitem [{\citenamefont {Glasser}, \citenamefont {Sundaresan},\ and\
  \citenamefont {Kevrekidis}(1998)}]{glasser98}%
  \BibitemOpen
  \bibfield  {author} {\bibinfo {author} {\bibfnamefont {B.~J.}\ \bibnamefont
  {Glasser}}, \bibinfo {author} {\bibfnamefont {S.}~\bibnamefont {Sundaresan}},
  \ and\ \bibinfo {author} {\bibfnamefont {I.~G.}\ \bibnamefont {Kevrekidis}},\
  }\bibfield  {title} {\enquote {\bibinfo {title} {From bubbles to clusters in
  fluidized beds},}\ }\href@noop {} {\bibfield  {journal} {\bibinfo  {journal}
  {Phys. Rev. Lett.}\ }\textbf {\bibinfo {volume} {81}},\ \bibinfo {pages}
  {1849--1852} (\bibinfo {year} {1998})}\BibitemShut {NoStop}%
\bibitem [{\citenamefont {Duru}\ and\ \citenamefont
  {Guazzelli}(2002)}]{duru02}%
  \BibitemOpen
  \bibfield  {author} {\bibinfo {author} {\bibfnamefont {P.}~\bibnamefont
  {Duru}}\ and\ \bibinfo {author} {\bibfnamefont {E.}~\bibnamefont
  {Guazzelli}},\ }\bibfield  {title} {\enquote {\bibinfo {title} {Experimental
  investigation on the secondary instability of liquid-fluidized beds and the
  formation of bubbles},}\ }\href@noop {} {\bibfield  {journal} {\bibinfo
  {journal} {Journal of Fluid Mechanics}\ }\textbf {\bibinfo {volume} {470}},\
  \bibinfo {pages} {359--382} (\bibinfo {year} {2002})}\BibitemShut {NoStop}%
\bibitem [{\citenamefont {Sundaresan}(2003)}]{sundaresan03}%
  \BibitemOpen
  \bibfield  {author} {\bibinfo {author} {\bibfnamefont {S.}~\bibnamefont
  {Sundaresan}},\ }\bibfield  {title} {\enquote {\bibinfo {title}
  {Instabilities in fluidized beds},}\ }\href@noop {} {\bibfield  {journal}
  {\bibinfo  {journal} {Annual Review of Fluid Mechanics}\ }\textbf {\bibinfo
  {volume} {35}},\ \bibinfo {pages} {63--88} (\bibinfo {year}
  {2003})}\BibitemShut {NoStop}%
\end{thebibliography}%

\end{document}